\documentclass[aps,prb,twocolumn,showpacs,preprintnumbers,amsmath,amssymb,superscriptaddress]{revtex4-2}
\usepackage{amsmath,amssymb,ascmac,fancybox}
\usepackage{booktabs}
\usepackage{color}
\usepackage{graphicx}
\usepackage{url}
\usepackage{siunitx}
\usepackage{bm}
\usepackage{physics}
\usepackage{mathtools}
\usepackage[colorlinks=true, allcolors=blue]{hyperref}
\usepackage[T1]{fontenc}

\newcommand{\ene}{\varepsilon}
\newcommand{\feq}{f^{\mathrm{eq}}}
\newcommand{\nneq}{n^{\mathrm{eq}}}
\newcommand{\kB}{k_{\mathrm{B}}}
\newcommand{\Te}{T_{\mathrm{e}}}
\newcommand{\betae}{\beta_{\mathrm{e}}}
\newcommand{\Tp}{T_{\mathrm{p}}}
\newcommand{\betap}{\beta_{\mathrm{p}}}

\newcommand{\RR}{\mathrm{R}}
\newcommand{\LL}{\mathrm{L}}
\newcommand{\gF}{g_{\mathrm{F}}}
\newcommand{\gB}{g_{\mathrm{B}}}
\newcommand{\NR}{N_{\RR}}
\newcommand{\NL}{N_{\LL}}
\newcommand{\sign}{\mathrm{sgn}\qty}

\newcommand{\cre}[1]{\hat{#1}^{\dagger}}
\newcommand{\ani}[1]{\hat{#1}}

\newcommand{\Wtot}{\mathcal{W}}

\begin{document}
\title{Theory of Seebeck ratchet in noncentrosymmetric electron-phonon coupled system}
\author{Yugo Onishi}
\affiliation{
Department of Applied Physics, University of Tokyo, 7-3-1 Hongo, Bunkyo-ku, Tokyo 113-8656
}

\author{Hiroki Isobe}
\affiliation{
Department of Applied Physics, University of Tokyo, 7-3-1 Hongo, Bunkyo-ku, Tokyo 113-8656
}

\author{Naoto Nagaosa}
\affiliation{
Center for Emergent Matter Science (CEMS), RIKEN, Wako 351-0198, Japan
}
\affiliation{
Department of Applied Physics, University of Tokyo, 7-3-1 Hongo, Bunkyo-ku, Tokyo 113-8656
}
\date{\today}

\begin{abstract}
    A thermoelectric effect of the noncentrosymmetric electron-phonon coupled system is proposed, in which a temperature difference between electrons and phonons, instead of a spatial temperature gradient, induces an electric current. This is a realization of the Seebeck ratchet in solids with the asymmetric energy dispersion of electrons, and the dynamical phonons act as the fluctuation of the scalar potential and induce the dc current. Possible realizations of this mechanism are also discussed.
\end{abstract}

\maketitle

\section{Introduction}
The thermoelectric effect is the effect of conversion between heat and electricity. Since they are promising for applications such as reuse of exhaust heat and energy harvesting, thermoelectric materials with high efficiency have been actively explored \cite{Dresselhaus2007,Mizuguchi2019,GutierrezMoreno2020}. The performances of thermoelectric materials are characterized by the conductivity $\sigma$, the thermal conductivity $\kappa$, and the Seebeck coefficient $S$, and the efficiency of the material is often measured by the dimensionless figure of merit $ZT=\sigma S^2 T/\kappa$ with the temperature $T$.

The thermoelectric effects most frequently discussed are the ones in metals or semiconductors, where electrons or holes serve as carriers of charge and heat. These systems are well described by the semiclassical Boltzmann theory and the quantum linear response theory \cite{Behnia2015, Luttinger1964, Kubo1957a, Ogata2019}. It is also well known that there exist several universal relations between thermoelectric coefficients, such as the Wiedemann-Franz law and the Mott formula. These relations impose some constraints on the thermoelectric effects of electronic origins in searching for efficient thermoelectric materials. In fact, the Wiedemann-Franz law shows that the conductivity and the thermal conductivity due to electrons are not independent, which makes it difficult to improve $ZT$. Under these constraints, various strategies to design efficient thermoelectric materials have been proposed, such as band engineering and low-dimensional materials \cite{Pei2012,Chen2003}.

While only the electrons are the carriers of charge in materials, other degrees of freedom such as phonons can be carriers of heat. Since phonons contribute to the heat current and the thermal conductivity, it is preferable to reduce the thermal conductivity due to phonons in order to improve $ZT$. Therefore, the most basic design strategy for thermoelectric materials was “phonon-glass, electron-crystal” \cite{Chen2003}. However, degrees of freedom such as phonons sometimes work positively for the thermoelectric effects. For example, in the phonon drag effect, phonons contribute to the Seebeck effect because electrons are dragged by the heat current of phonons. The phonon drag effect can be also described by the Boltzmann theory and the linear response theory. This phonon drag effect can lead to a large Seebeck effect at low temperatures in, e.g., doped semiconductors \cite{Herring1954,Mahan2014} and, was proposed recently as an origin of the large Seebeck effect at low temperature in FeSb$_2$ \cite{Matsuura2019a}. Since thermoelectric effects using degrees of freedom other than electrons can avoid restrictions such as Wiedemann-Franz law, they are one of the promising ways to achieve high-performance thermoelectric effects. Therefore, how degrees of freedom other than electrons can contribute to thermoelectric effects is an important issue in terms of applications. 

From this viewpoint, we explore the possible role of lower symmetry of the electron-phonon coupled system as a mechanism of the thermoelectric effect. Especially, nonreciprocal charge transport in noncentrosymmetric systems is a subject of recent intensive interest \cite{Tokura2018}. In addition to the broken inversion symmetry $\cal{P}$, the time-reversal symmetry $\cal{T}$ is often required to be broken due to the magnetic ordering or external magnetic field for the directional transport. This can be understood from the energy dispersion of the band structure, which satisfies $\ene_\uparrow(\vb{k}) = \ene_\downarrow(-\vb{k})$ when $\cal{T}$ is preserved. 
Hence the nonreciprocal charge transport is represented by the formula \cite{Tokura2018} 
\begin{equation}
J = \sigma_2  E^2 
\label{eq:nonlinear}
\end{equation}
where the nonlinear conductivity $\sigma_2$ is linear in the magnetic field $B$ for small $B$.
This indicates that the diode effect is switchable by the magnetic field, although the rectification effect is still very small.

In the presence of electron-phonon interaction, the phonons act as the fluctuation of the scalar potential and the consequent internal electric field $E^{\mathrm{ph}}$. When the external electric field $E$ in Eq.(\ref{eq:nonlinear}) is naively replaced by $E^{\mathrm{ph}}$, a dc current $J$ appears in equilibrium. Of course, this is wrong since there is no dc current with dissipation in thermal equilibrium, and a nonequilibrium steady state with energy flow is required to support the dc current. This situation is analogous to a Brownian ratchet, where the Brownian motion with different temperatures between two systems induces the rotational motion in one direction \cite{Reimann2002}. 

In this work, we propose a thermoelectric effect with the asymmetric electron dispersion driven by the phonons. Assuming that the electron temperature $\Te$ and that of phonons $\Tp$ are different, we find that the correct formula is 
\begin{equation}
J \propto  \expval{(E^{\mathrm{ph}})^2}_{\Tp} - \expval{(E^{\mathrm{ph}})^2}_{\Te}.   
\label{eq:dc}
\end{equation}
where the angle brackets $\expval{\dots}_T$ represent the thermal average with temperature $T$.
While the heat current flows spatially in the conventional thermoelectric effects, the heat current flows between phonons and electrons in the effect proposed here. This is the crucial difference from previous studies. For example, in \cite{Wurger2021} a mechanism similar to the ratchet is also proposed, but the mechanism is based on the spatial gradient of temperature in contrast to our proposal.

Our mechanism is similar to the one proposed for the magnon-driven spin-Seebeck effect \cite{Xiao2010a}. In the magnon-driven spin-Seebeck effect, the temperature difference between electrons and magnons induces the spin current. The results of the present work show that a similar mechanism induces a finite electric current in an electron-phonon coupled system.
The mechanism proposed by Bosisio {\it et al.} \cite{Bosisio2016} is closely related to the present work. While they discussed the behavior of the localized electrons in the presence of the phonons when there is a temperature difference between electrons and phonons, we consider itinerant electrons and thus do analysis in $k$-space. 
It should be also noted that the inversion symmetry breaking (or, equivalently, reflection symmetry breaking in one-dimensional systems) is essential to realize the thermoelectric effects induced by the temperature difference between electrons and phonons. While in the model proposed by Bosisio {\it et al.} the asymmetric contacts between the system and the electrodes break the inversion symmetry, the dispersion of the electrons breaks the symmetry in our model. In that sense, the effect proposed in the present work is a bulk effect in contrast to the proposal by Bosisio {\it et al.} Another closely related work is done by Budkin {\it et al.} \cite{Budkin2020}. They proposed that the shift of the position of electrons is induced by phonons when the effective temperatures for electrons and phonons are different. The shift is described by the Berry connection as in the case of shift current \cite{VonBaltz1981,Morimoto2016}. However, this effect is also different from our proposal because the effect proposed in \cite{Budkin2020} is an interband effect while the effect proposed in the present work is an intraband effect.

This paper is organized as follows. In Sec.\ref{sec:model}, we present a model for an electron-phonon coupled system considered in the present work. In Sec.\ref{sec:calc_results}, we solve the model perturbatively and calculate an electric current and heat current induced by the temperature difference between electrons and phonons. We also calculate quantities which correspond to the power factor and the figure of merit in conventional thermoelectric materials. Sec.\ref{sec:Discussion} is devoted to discussions and conclusions. We also discuss the possible realization of the proposed mechanism such as excitation by light and a metal-insulator junction.

\section{Model} \label{sec:model}
\begin{figure}[htbp]
    \centering
    \includegraphics[width=0.8\columnwidth]{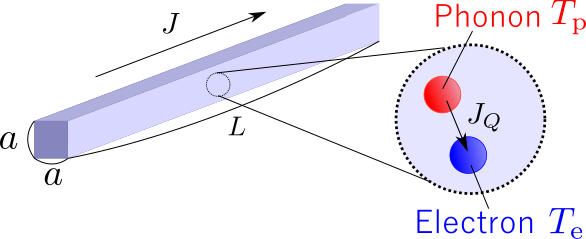}
    \caption{One-dimensional system considered in the present work. The length of the system is $L$ and the cross-section is $a\times a$. The electric current flows along the wire, while the heat current flows between phonons with temperature $\Tp$ and electrons with temperature $\Te$.}
    \label{fig:geometry}
\end{figure}

Let us consider a model which consists of spinless electrons with an asymmetric dispersion and phonons in one dimension (Fig.\ref{fig:geometry}). We assume that electrons have dispersion $\ene_k$ and phonons have dispersion $\omega_q$. The behavior of electrons can be described with the Boltzmann equation:
\begin{align}
    \pdv{f}{t} + \dot{r}\pdv{f}{r} + \dot{k}\pdv{f}{k} = \qty(\pdv{f}{t})_{\mathrm{scatt}} \label{eq:Boltzmann}
\end{align}
Here, $f(r,k,t)$ is the distribution function of electrons, and the right-hand side describes the scattering of electrons. We assume that the scattering term is given by two terms: one corresponds to the scattering due to the electron-phonon interaction and the other corresponds to the other sources of scattering such as impurities. They can be expressed as follows.
\begin{align}
    &\qty(\pdv{f_k}{t})_{\mathrm{scatt}} = -\frac{f_k-\feq_k}{\tau} + \qty(\pdv{f_k}{t})_{\mathrm{phonon}} \label{eq:scattering}\\
    &\qty(\pdv{f_k}{t})_{\mathrm{phonon}} = \sum_{k'} \qty(\Wtot_{k'\to k} - \Wtot_{k\to k'}) \label{eq:el-ph_scattering1}\\
    &\Wtot_{k+q \to k}[f_k, n_q] = \frac{2\pi}{\hbar} \frac{\abs{g_q}^2}{V} f_{k+q}\qty(1-f_k) \nonumber\\ 
    &\times\qty[ n_q \delta(\ene_{k+q} - \ene_k+\hbar\omega_q) + \qty(n_q+1)\delta(\ene_{k+q}-\ene_k-\hbar\omega_q) ] \label{eq:el-ph_scattering2}
\end{align}
where $f_k$ is shorthand for $f(r,k,t)$, and since we are interested only in the spatially uniform steady state, we omit $r$ and $t$ dependencies of $f(r,k,t)$. $\feq_k = (1+e^{\beta_e(\ene_k-\mu)})^{-1}$ is the equilibrium distribution function of electron with temperature $\Te = 1/\kB\betae$ and chemical potential $\mu$. 
In Eq.\eqref{eq:scattering}, we use the relaxation time approximation, where  $\tau$ accounts for scatterings other than the electron-phonon interaction.
Here we assume that $\tau$ is independent of $k$ for simplicity.
The expression of electron-phonon scattering term is given in Eq.\eqref{eq:el-ph_scattering1}. $\Wtot_{k'\to k}$ is the transition rate of an electron scattered from state $k'$ to $k$, and is given as Eq.\eqref{eq:el-ph_scattering2} by the Fermi's golden rule. $g_q$ is the coupling constant between phonons and electrons, $n_q = n(r,q,t)$ is phonon distribution function, and $V=La^2$ is the volume of the system with length $L$ and cross-section $a^2$ (Fig.\ref{fig:geometry}). We do not consider $r$ and $t$ dependencies of $n(r,q,t)$ as in the case of $f(r,k,t)$.

If electrons and phonon are in equilibrium with temperature $T = 1/\kB\beta$ and chemical potential $\mu$, i.e., 
\begin{align}
    f(r,k,t) &= \frac{1}{e^{\beta(\ene_k-\mu)}+1} \\
    n(r,q,t) &= \frac{1}{e^{\beta\hbar\omega_q}-1}
\end{align}
then we can confirm $\Wtot_{k'\to k} = \Wtot_{k\to k'}$. This equality comes from the following identity which holds only in equilibrium:
\begin{align}
    & f_{k'}(1-f_k) n_q \delta(\ene_{k'}-\ene_k+\omega_q) \nonumber \\
    & = f_{k}(1-f_{k'})(1+n_q) \delta(\ene_{k'}-\ene_k+\omega_q) \label{eq:detailed_balance}
\end{align}
Therefore, in equilibrium, electron-phonon scattering term, Eq.\eqref{eq:el-ph_scattering1}, is zero.

\begin{figure}[htbp]    
    \centering
    \includegraphics[width=0.7\columnwidth]{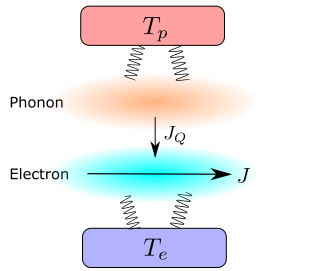}
    \caption{The thermoelectric effect proposed in this paper. Instead of a spatial gradient of temperature, a temperature difference between phonons and electrons induces an electric current. The direction in which the electric current flows is set by the inversion symmetry breaking of the electron system.}
    \label{fig:schematics}
\end{figure}

In this paper, we discuss an unconventional thermoelectric effect of the system described above. 
Conventionally, the source of the thermoelectric effects is a spatial gradient of temperature. 
However, we consider another type of thermoelectric effect. Instead of a spatial temperature gradient, we consider a temperature difference between electrons and phonons as schematically shown in Fig.\ref{fig:schematics} (For the realization of such a system, see Sec.\ref{sec:Discussion}). As we confirmed above, if electrons and phonons feel the same temperature, the system relaxes into the equilibrium state and no currents flow. However, we show that electric currents can occur when electrons and phonons feel different temperatures. 

\section{Calculation and Results} \label{sec:calc_results}
\subsection{Solution of the Boltzmann equation}
In this section, we solve the Boltzmann equation Eq.\eqref{eq:Boltzmann} up to $\order{\abs{g_q}^2}$, when electrons and phonons feel different temperatures, $\Te$ for electrons and $\Tp$ for phonons. 

Before proceeding to detailed calculations, we remark on the phonon scattering term $\Wtot_{k'\to k}$. What we are interested in here is the effect due to temperature difference $\Tp-\Te$. In other words, we are interested in the effect by $\delta \nneq_q = \nneq_q(\Tp)-\nneq_q(\Te)$, where $\nneq_q(T)$ is the equilibrium distribution function at temperature $T$ for phonon with wavevector $q$. However, $\Wtot_{k'\to k}$ includes contribution from the deviation of $f_k$ from the equilibrium distribution at $\Te$, $\delta f_k = f_k-\feq_k$.
To clearly distinguish these two contributions, we rewrite $\Wtot_{k'\to k}$ by expanding $\Wtot_{k'\to k}$ with respect to the deviation from the equilibrium distribution, $\delta f_k$ and $\delta\nneq_q$. 
\begin{align}
    \Wtot_{k'\to k}[f_k, \nneq_q(\Tp)] &= W^{(0)}_{k'\to k} + \tilde{W}_{k'\to k}[\delta f_k] + W_{k'\to k}[\delta\nneq_q] \\
	W^{(0)}_{k'\to k} &= \Wtot_{k'\to k}[\feq_k(\Te), \nneq_q(\Te)]
\end{align} 
Here we neglect higher order terms such as $\delta f_k \delta\nneq_q$. To justify this expansion, $\delta f_k$ and $\delta \nneq_q$ need to be sufficiently small. This condition is discussed at the end of this section.
The first term is $\Wtot_{k'\to k}$ when $f_k=\feq_k(\Te)$ and $n_q = \nneq_q(\Te)$. Here, we explicitly write $\Te$ dependence of $\feq_k$ to emphasize that $\feq_k$ is the equilibrium distribution for $T=\Te$. Since $W^{(0)}_{k'\to k}=W^{(0)}_{k\to k'}$, $W^{(0)}_{k'\to k}$ does not affect the Boltzmann equation, as mentioned at the end of the last section. The second term is the contribution from $\delta f_k$ and the third is the contribution from $\delta\nneq_q$. The second term can be incorporated into the relaxation time $\tau$ in Eq.\eqref{eq:scattering}. 
The third term $W_{k'\to k}$ can be obtained by expanding $\Wtot_{k'\to k}$ with respect to $\delta\nneq_q$.
\begin{align}
    &W_{k+q \to k} = \frac{2\pi}{\hbar} \frac{\abs{g_q}^2}{V} \feq_{k+q}\qty(1-\feq_k) \nonumber\\ 
    &\times\qty[\delta(\ene_{k+q} - \ene_k+\hbar\omega_q) + \delta(\ene_{k+q}-\ene_k-\hbar\omega_q) ]\delta \nneq_q \label{eq:el-ph_scattering3}
\end{align}

Let us now turn to solving the Boltzmann equation. We assume that there is no spatial variation of temperature and that there is no electric field. In this case, the Boltzmann equation for steady state is 
\begin{align}
    -\frac{f_k-\feq_k}{\tau} + \sum_{k'} \qty(W_{k'\to k} - W_{k\to k'}) = 0 \label{eq:Boltzmann_ss}
\end{align}
Here, $f_k = f(r, k, t)$ does not depend on $r$ or $t$. Therefore, we consider only $k$-dependence of the distribution function $f_k$. This equation can be solved as
\begin{align}
    f_k &= \feq_k + \tau\sum_{k'}\qty(W_{k'\to k} - W_{k\to k'}) \label{eq:solution1}
\end{align}

\begin{figure}[htbp]
    \centering
    \includegraphics[width=0.8\columnwidth]{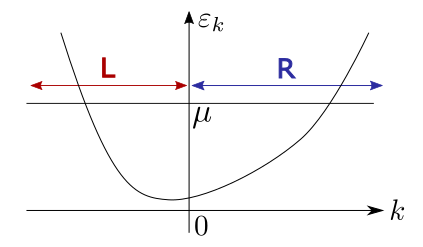}
    \caption{Asymmetric one-dimensional band structure. We consider the case that there are two Fermi points. Correspondingly we separate the Brillouin zone into two parts, $\RR$ and $\LL$.}
    \label{fig:band_structure}
\end{figure}

For analytic calculations, we further make three simplifications. First, we consider a one-dimensional dispersion, i.e., $a$ is small enough so that $\hbar^2/2ma^2 \gg \mu$ where $\mu$ is chemical potential and $m$ is the effective mass of an electron. Second, we ignore the phonon dispersion and set $\omega_q = \omega$(constant). Third, we simplify the $q$-dependence of the coupling constant $g_q$ as 
\begin{align}
    g_q &= \begin{cases}
        \gF \quad \text{when both $k, k+q$ are on the same side, $\RR$ or $\LL$} \\
        \gB \quad \text{when $k, k+q$ are on different side, $\RR$ and $\LL$}
    \end{cases} \label{eq:gq_dependence_assumption}
\end{align}
Here, R and L denote regions in the Brillouin zone of electrons near two Fermi points (see Fig.\ref{fig:band_structure}). Corresponding to R and L, the density of states of electrons $N(\xi)$ are separated into two parts: 
\begin{align}
    N(\xi) &= \NR(\xi) + \NL(\xi) \\
    \NR(\xi) &= \frac{1}{V}\sum_{k\in \RR} \delta(\xi_k-\xi) \\
    \NL(\xi) &= \frac{1}{V}\sum_{k\in \LL} \delta(\xi_k-\xi)
\end{align}
where $\xi_k = \ene_k-\mu$.

Under these two assumption, we can explicitly calculate the $k'$-summation in Eq.\eqref{eq:solution1} as follows. With the functions 
\begin{align}
    \feq(\xi) &= \frac{1}{e^{\beta_e\xi}+1} \\
    F(\xi_1, \xi_2) &= \feq(\xi_1)(1-\feq(\xi_2)) \\
    \nneq(\omega, \Tp)&=\frac{1}{e^{\betap\hbar\omega}-1}
\end{align}
we can calculate the $k'$ summation (see Appendix \ref{Ap:sol_BE} for details) to obtain
\begin{widetext}
\begin{subequations}
    \begin{align}
        \text{$k\in R$:} \quad &\sum_{k'}\qty(W_{k'\to k} - W_{k\to k'}) \nonumber \\
        &= \frac{2\pi}{\hbar} \qty(\abs{\gB}^2\NL(\xi_k-\hbar\omega) + \abs{\gF}^2\NR(\xi_k-\hbar\omega)) \qty[F(\xi_k-\hbar\omega,\xi_k) -F(\xi_k,\xi_k-\hbar\omega)]\delta\nneq(\omega)  \nonumber \\
        & \quad + \frac{2\pi}{\hbar} \qty(\abs{\gB}^2\NL(\xi_k+\hbar\omega) + \abs{\gF}^2\NR(\xi_k+\hbar\omega)) \qty[F(\xi_k+\hbar\omega,\xi_k)-F(\xi_k,\xi_k+\hbar\omega)] \delta \nneq(\omega) \label{eq:k_R_Wsum}\\
        \text{$k\in L$:} \quad &\sum_{k'}\qty(W_{k'\to k} - W_{k\to k'}) \nonumber \\
        &= \frac{2\pi}{\hbar} \qty(\abs{\gF}^2\NL(\xi_k-\hbar\omega) + \abs{\gB}^2\NR(\xi_k-\hbar\omega)) \qty[F(\xi_k-\hbar\omega,\xi_k)-F(\xi_k,\xi_k-\hbar\omega)] \delta\nneq(\omega) \nonumber \\
        & \quad + \frac{2\pi}{\hbar} \qty(\abs{\gF}^2\NL(\xi_k+\hbar\omega) + \abs{\gB}^2\NR(\xi_k+\hbar\omega)) \qty[F(\xi_k+\hbar\omega,\xi_k)-F(\xi_k,\xi_k+\hbar\omega)]\delta\nneq(\omega) \label{eq:k_L_Wsum} 
    \end{align} \label{eq:Wsum}
\end{subequations}
Here, an abbreviation $\delta \nneq(\omega) = \nneq(\omega, \Tp)-\nneq(\omega, \Te)
$ is used.
We calculate the electric current and the heat current from phonons to electrons using these results in the next section. We note here that the following calculation is valid only when the correction to the distribution function due to scattering is sufficiently small, i.e.,
\begin{align}
    \abs{\tau \sum_{k'}(W_{k'\to k}-W_{k\to k'})} \ll 1 \label{eq:condition_valid_1}
\end{align}
This condition is satisfied when $\abs{\Tp-\Te}$ is sufficiently small as derived in the following. 
To transform this condition into a simpler form, the following identity is useful.
\begin{align}
    F(\xi-\hbar\omega,\xi) -F(\xi,\xi-\hbar\omega) &= \frac{F(\xi, \xi-\hbar\omega)}{\nneq(\omega, \Te)} 
\end{align}
By applying this identity to Eqs.\eqref{eq:Wsum}, 
we can show that the inequality \eqref{eq:condition_valid_1} is satisfied when the following inequality is satisfied.
\begin{align}
    2\pi\tau\betae\omega\abs{g_{\mathrm{F/B}}}^2 N_{\mathrm{R/L}} (\nneq(\omega,\Te)+1) \abs{\frac{\Tp-\Te}{\Te}}\ll 1 \label{eq:condition_valid}
\end{align}
where $\mathrm{F/B}$ and $\mathrm{R/L}$ are arbitrarily chosen. If the scattering is mainly due to electron-phonon interaction, the order of $\hbar/2\pi\tau_{k}$ for an electron with $\xi_k\simeq 0$ can be estimated as \cite{Mahan2000}
\begin{align}
    \frac{\hbar}{2\pi\tau_k} &= \sum_{q}\frac{\abs{g_q}^2}{V} \qty[ (n_q+1-f_{k+q}) \delta(\ene_{k+q} - \ene_k+\hbar\omega_q) + \qty(n_q+f_{k+q})\delta(\ene_{k+q}-\ene_k-\hbar\omega_q)] \nonumber \\
    &\sim \abs{g_{\mathrm{F/B}}}^2 (\nneq(\omega, \Tp)+1)N_{\mathrm{R/L}} \label{eq:tau_estimate}
\end{align}
and therefore the condition Eq.\eqref{eq:condition_valid} becomes much simpler form in this case:
\begin{align}
    \frac{\hbar\omega}{\kB T_e} \abs{\frac{\Tp-\Te}{\Te}}\ll 1 \label{eq:condition_valid_el-ph}
\end{align}

\subsection{Calculation of the electric current} \label{sec:calc_J}
The electric current $J$ is given by 
\begin{align}
    J &= \frac{e}{L}\sum_{k} \frac{1}{\hbar}\pdv{\xi_k}{k} f_k %
    = \frac{e\tau}{\hbar L}\sum_{k}\pdv{\xi_k}{k}\sum_{k'}\qty(W_{k'\to k} - W_{k\to k'}) \label{eq:current}
\end{align}
where $e(<0)$ is the charge of electrons.  
Here we used the fact that in equilibrium $J = e/(\hbar L) \sum_k \pdv*{\xi_k}{k}\feq_k = 0$.

We can take $k$-summation in Eq.\eqref{eq:current} explicitly. 
After some algebra, we obtain
\begin{align}
    J = \frac{e\tau}{\hbar}\int_{-\infty}^{\infty} \frac{\dd{\xi}}{\hbar} &(-1)^{\RR} \qty(\abs{\gF}^2-\abs{\gB}^2) \qty((\NL(\xi-\hbar\omega)-\NL(\xi)) - (\NR(\xi-\hbar\omega)-\NR(\xi))) \nonumber \\
    &\times \qty[F(\xi-\hbar\omega,\xi)-F(\xi,\xi-\hbar\omega)]\delta\nneq(\omega)  \label{eq:Wsum1}
\end{align}
where the factor $(-1)^{\RR/\LL}$ is defined as $(-1)^{\RR/\LL}=\sign(\eval{\pdv{\xi}{k}}_{\RR/\LL})$, and we have used the assumption $(-1)^\LL=-(-1)^\RR$. 

We can further simplify this expression as
\begin{align}
    J &= (-1)^{\RR} e\tau\omega^2 \qty(\abs{\gF}^2-\abs{\gB}^2) \qty(\NR'(0)-\NL'(0)) %
    \qty[\nneq(\omega, \Tp)-\nneq(\omega, \Te)] \label{eq:current_by_T}
\end{align}
Here, $\NR'(\xi)$ is the $\xi$-derivative of $\NR(\xi)$ and similar for $\NL'(\xi)$.
For details of the derivation of Eqs.\eqref{eq:Wsum1},\eqref{eq:current_by_T}, see Appendix \ref{Ap:calc_J_by_dT}. 
From Eq.\eqref{eq:current_by_T}, we conclude that a temperature difference between phonons and electrons can induce electric currents if $\abs{\gF}^2\neq \abs{\gB}^2$ and $\NL'(0)\neq\NR'(0)$.

\subsection{Calculation of the heat current}
We can also calculate heat current $J_Q$ from phonons to electrons as follows. $J_Q$ can be defined as 
\begin{align}
    J_Q &= \sum_k \xi_k \qty(\pdv{f_k}{t})_{\mathrm{phonon}} 
    = \sum_{k}\xi_k\sum_{k'}(\Wtot_{k'\to k}-\Wtot_{k\to k'}) \label{eq:heat_current_def1}
\end{align}
This expression means that the change of energy of electrons due to phonon scattering is equal to the heat current from phonons to electrons. Note that we use here $\Wtot_{k'\to k}$ instead of $W_{k'\to k}$.

Up to $\order{\abs{g_q}^2}$, $\Wtot_{k'\to k}-\Wtot_{k\to k'} = W_{k'\to k}-W_{k\to k'}$ in the absence of an electric field. By using Eqs.\eqref{eq:Wsum},\eqref{eq:heat_current_def1}, we obtain 
\begin{align}
    J_Q 
    &= 2\pi V\hbar \omega^2 \qty[\abs{\gF}^2(\NL(0)^2+\NR(0)^2) + 2\abs{\gB}^2 \NR(0)\NL(0)]\qty[\nneq(\omega,\Tp)-\nneq(\omega, \Te)] \label{eq:heat_current_by_T}
\end{align}
where we have used an approximation $\NR(\xi)\simeq\NR(0)$ and $\NL(\xi)\simeq\NL(0)$ (see Appendix \ref{Ap:calc_J_by_dT} for details). Since $J_Q$ represents the heat current between electrons and phonons, it is proportional to the volume $V=La^2$.
In particular, the ratio between $J$ and $J_Q$ is 
\begin{align}
    \frac{J}{J_Q} &= (-1)^\RR \frac{e\tau}{2\pi V} \frac{\qty(\abs{\gF}^2-\abs{\gB}^2) \qty(\NR'(0)-\NL'(0))}{\abs{\gF}^2(\NL(0)^2+\NR(0)^2) + 2\abs{\gB}^2 \NR(0)\NL(0)}
\end{align}
This ratio is independent of temperature. 

\subsection{Linear response coefficients}
We can now calculate linear response coefficients. We define $2\times 2$ matrix $L$ as:
\begin{subequations}
    \begin{align}
        j &= L_{11} E + L_{12} \qty(-\frac{\Delta T}{T}) \\
        q &= L_{21} E + L_{22} \qty(-\frac{\Delta T}{T})
    \end{align}     
\end{subequations}
where $E$ is an external electric field, $T=\Te$ is the temperature of electrons, and $\Delta T$ is the temperature difference between phonons and electrons, $\Delta T = \Te -\Tp$. $j=J/a^2$ is an electric current density, and $q=J_Q/V$ is a heat current per unit volume. Note that $J_Q$ represents the total heat current between electrons and phonons, and thus it is proportional to the volume $V$.

$L_{12}$ and $L_{22}$ can be calculated from Eq.\eqref{eq:current_by_T} and \eqref{eq:heat_current_by_T} as 
\begin{align}
    L_{12} &= (-1)^{\RR} \frac{e\tau\hbar\omega^3}{a^2 \kB T} \qty(\abs{\gF}^2-\abs{\gB}^2) \qty(\NR'(0)-\NL'(0)) \nneq(\omega)(\nneq(\omega)+1) \\
    L_{22} &= \frac{2\pi \hbar^2 \omega^3}{\kB T} \qty[\abs{\gF}^2(\NL(0)^2+\NR(0)^2) + 2\abs{\gB}^2 \NR(0)\NL(0)] \nneq(\omega)\qty(\nneq(\omega)+1)
\end{align}

We can also calculate $L_{11}$ and $L_{21}$ by solving the Boltzmann equation in the presence of an electric field.
\begin{align}
    \frac{eE}{\hbar}\pdv{f_k}{k} &= -\frac{f_k-\feq_k}{\tau} + \sum_{k'}\qty(W_{k'\to k}- W_{k\to k'})
\end{align}
By solving the equation up to $\order{E\abs{g_q}^2}$, we can calculate $J$ and $J_Q$ (see Appendix.\ref{Ap:calc_linres} for details) and we obtain 
\begin{align}
    L_{11} &= \frac{e^2\tau}{4\pi^2\hbar^2 a^4}\qty(\frac{1}{\NR(0)}+\frac{1}{\NL(0)}) 
    \label{eq:L11} \\
    L_{21} &= (-1)^\RR\frac{e\tau\hbar\omega^3}{a^2 \kB T}(\abs{\gF}^2-\abs{\gB}^2)(\NR'(0)-\NL'(0))\nneq(\omega,\Te)(\nneq(\omega,\Te)+1) = L_{12} \label{eq:L21}
\end{align} 
Since $L_{12}=L_{21}$, the Onsager's reciprocal relationship holds. From these expressions, we define the following quantities: $\sigma$, $\lambda$, $p$, and $ZT$.
\begin{subequations}
    \begin{align}
        \sigma &:= L_{11} = \frac{e^2\tau}{4\pi^2\hbar^2 a^4}\qty(\frac{1}{\NR(0)}+\frac{1}{\NL(0)}) \\
        \lambda &:= \qty(\frac{q}{\Delta T})_{j=0} = \frac{L_{11}L_{22}-L_{12}L_{21}}{TL_{11}} \simeq \frac{L_{22}}{T} \nonumber \\
        &= \frac{2\pi \hbar^2\omega^3}{\kB T^2} \qty[\abs{\gF}^2(\NL(0)^2+\NR(0)^2) + 2\abs{\gB}^2 \NR(0)\NL(0)] \nneq(\omega)\qty(\nneq(\omega)+1) \\
        p &:= \frac{L_{12}^2}{T^2L_{11}} \nonumber \\
        &= \frac{4\pi^2 \hbar^4\omega^6 \tau}{\kB^2 T^4}\frac{1}{\NR(0)^{-1}+\NL(0)^{-1}}\qty[(\abs{\gF}^2-\abs{\gB}^2)(\NR'(0)-\NL'(0)) \nneq(\omega)(\nneq(\omega)+1)]^2\\
        ZT &:= \frac{L_{12}^2}{L_{11}L_{22}-L_{12}L_{21}} \simeq \frac{L_{12}^2}{L_{11} L_{22}} \nonumber \\
        &= \frac{2\pi\tau\hbar^2\omega^3}{\kB T} \frac{\qty{(\abs{\gF}^2-\abs{\gB}^2)(\NR'(0)-\NL'(0))}^2}{(\NR(0)^{-1}+\NL(0)^{-1}) \qty[\abs{\gF}^2(\NL(0)^2+\NR(0)^2) + 2\abs{\gB}^2 \NR(0)\NL(0)]}\nneq(\omega)(\nneq(\omega)+1) \label{eq:ZT}
    \end{align} \label{eq:linear_response_coefficients}
\end{subequations}
\end{widetext}
Here, $\sigma$ represents the usual conductivity. $\lambda$ represents the heat current response to the temperature difference $\Delta T$ under the condition $j=0$. This is an analogous quantity to the thermal conductivity in thermoelectric materials. $p$ is an analogous quantity to power factor for thermoelectric materials, and $p(\Delta T)^2/4$ is the maximum of power output per unit volume for a given $\Delta T$ \cite{Benenti2017}. $ZT$ is the figure of merit which represents the efficiency of the thermoelectric system \cite{Benenti2017}. Note, however, that $\lambda$ and $p$ are conceptually different from the thermal conductivity and power factor for thermoelectric materials. In fact, their dimensions are different from those of thermal conductivity and power factor. In the following, we call $\lambda$ heat flow rate and $p$ power density.

\section{Discussion} \label{sec:Discussion}
What we calculated in the previous section is the electric current induced by the nonequilibrium distribution of phonons. As a similar phenomenon, the phonon drag effect on the Seebeck coefficient is well known and extensively investigated \cite{Herring1954,Zhou2015,Matsuura2019a}. Although these two effects might look quite similar, their origins are qualitatively different. In phonon drag, the distribution of phonons which affects the electric current has nonzero net velocity since phonons carry the heat current. These phonons with finite velocity “drag” an electric current. On the other hand, in our calculation, we assume dispersionless phonons. Therefore, their group velocity is zero, resulting in no net velocity. 
Instead, the phonon system excites electron-hole pairs with finite velocity.
This picture become clearer when Eq.\eqref{eq:Wsum1} is transformed into the following form.
\begin{align}
    &J = \frac{e\tau}{\hbar}\int_{-\infty}^{\infty} \frac{\dd{\xi}}{\hbar} (-1)^{\RR} \qty(\abs{\gF}^2-\abs{\gB}^2) \nonumber \\ 
    &\times \qty((\NL(\xi-\hbar\omega)-\NL(\xi)) - (\NR(\xi-\hbar\omega)-\NR(\xi))) \nonumber \\
    &\times \qty[F(\xi-\hbar\omega,\xi) \nneq(\omega,\Tp)-F(\xi,\xi-\hbar\omega)\qty(\nneq(\omega,\Tp)+1)] \label{eq:el-hole_pair}
\end{align}
This form can also be derived directly by using $\Wtot_{k'\to k}$ instead of $W_{k'\to k}$. The function $F(\xi-\hbar\omega, \xi)$ in Eq.\eqref{eq:el-hole_pair} is nothing but the probability for electron-hole pair, i.e., an electron with energy $\xi-\hbar\omega$ and a hole with energy $-\xi$, to exist. Therefore, Eq.\eqref{eq:el-hole_pair} can be understood as an electron-hole pair excitation (relaxation) by a phonon absorption (emission). 
In equilibrium, the excitations and relaxations of electron-hole pairs balance and therefore no net current flows, but if the phonons and the electrons have different temperatures, this balance breaks and an electric current flows. 

Another way to interpret this effect is to view the fluctuation of phonons as an effective electric field fluctuation. The electron-phonon interaction term we consider originally comes from the following Hamiltonian:
\begin{align}
    \hat{H}_{\mathrm{e-ph}} &= \frac{1}{\sqrt{V}}\sum_{\vec{k}} g_{\vec{q}}\hat{\rho}_{\vec{q}}(\ani{a}_{\vec{q}} + \cre{a}_{-\vec{q}})
\end{align}
where $\hat{\rho}_{\vec{q}}$ is the Fourier component of electron density, and $\ani{a}_{\vec{q}}$($\cre{a}_{\vec{q}}$) is the annihilation(creation) operator of phonon with wavevector $\vec{q}$. Therefore, $g_{\vec{q}}(\ani{a}_{\vec{q}} + \cre{a}_{-\vec{q}})$ can be interpreted as an effective electric field $E_{\vec{q}}^{\mathrm{ph}}$. On the other hand, the electric current $J$ under electric field $E$ can be expanded with respect to $E$ as
\begin{align}
    J &= \sigma_1 E + \sigma_2 E^2 + \dots
\end{align}
The second term represents the second order response. 
The expectation value $\expval{E_{\vec{q}}^{\mathrm{ph}}}$ is always zero and thus no current flows due to linear response. However, its fluctuation, $\expval{(E_{\vec{q}}^{\mathrm{ph}})^2} \sim \abs{g_q}^2\expval{\cre{a}_{\vec{q}}\ani{a}_{\vec{q}}}$, is nonzero and can induce an electric current through the second order response. The corresponding term in Eq.\eqref{eq:current_by_T} is the term proportional to $\nneq(\omega, \Tp)$. Eq.\eqref{eq:current_by_T} also shows that this current is cancelled by the current proportional to $\nneq(\omega, \Te)$, as in Eq.\eqref{eq:dc}. Therefore, no net current flows in equilibrium, but if the phonons and electrons have different temperatures, the net current can be finite.

As we mentioned in Sec.\ref{sec:calc_J}, our calculation shows that it is necessary to have $\abs{\gF}^2\neq \abs{\gB}^2$ and $\NL'(0)\neq\NR'(0)$ to induce finite a thermoelectric current. The first condition requires $q$-dependence of electron-phonon interaction parameter, $g_q$. The second condition requires that an asymmetric band structure of electrons is necessary.
To realize an asymmetric band structure in one-dimensional systems, both time reversal symmetry and spatial inversion symmetry need to be broken. Otherwise, wavevector $k$ and $-k$ can be related to each other through the symmetries and the two Fermi points $\LL$ and $\RR$ become equivalent, resulting in $\NL'(0) = \NR'(0)$. More physically, these symmetry breakings allow us to set a spatial direction of the materials. When there is a temperature difference between electrons and phonons, an electric current can flow in the direction. Because we now consider setting the spatial direction through the band dispersion of the electron, we need to break time reversal symmetry, but if the inversion symmetry is broken in other ways, e.g., asymmetric contacts with leads \cite{Bosisio2016}, the time reversal symmetry breaking is not necessary.

From  Eq.\eqref{eq:current_by_T} and Eq.\eqref{eq:heat_current_by_T}, the order of magnitude estimates of the current densities $j=J/a^2$ and the heat current per unit volume $q=J_Q/V$ are given by
\begin{align}
    j &\sim \frac{e\tau\omega^2}{a^2} (\Delta\abs{g}^2) (\Delta N'(0)) (\nneq(\omega,\Tp)-\nneq(\omega, \Te)) \\
    q &\sim \hbar\omega^2 \abs{g}^2 N(0)^2 (\nneq(\omega,\Tp)-\nneq(\omega, \Te))
\end{align}
where $\Delta \abs{g}^2 = \abs{\gF}^2-\abs{\gB}^2$, $\Delta N'(0)=\NR'(0)-\NL'(0), g\sim g_{\mathrm{F/B}}$, $N(0)\sim N_{\mathrm{R/L}}$.

To estimate the order of currents, we consider GeTe doped with ferromagnetic Mn impurities. GeTe is a noncentrosymmetric semiconductor and nonreciprocal charge transport under magnetic field is experimentally observed at room temperature \cite{Li2021}. By doping Mn, it also breaks time reversal symmetry. Therefore, it is a promising material to realize the proposed effect. We assume here that a temperature difference between electrons and phonons in this system is made by, for example, laser heating (for the realization of the temperature difference, see later discussions). Although Mn-doped GeTe is not a one-dimensional system, we roughly estimate the order of magnitude from our one-dimensional calculation.

We estimate the parameters for Mn-doped GeTe by the ones for GeTe \cite{Askarpour2019, Li2021}: $\hbar\omega \sim \SI{10}{meV}$, $\tau\sim\SI{1e-13}{s}$, $N(0)\sim\ene_F^{-1}a^{-3}$, $N'(0)\sim\ene_F^{-2}a^{-3}$, Fermi energy $\ene_F\sim\SI{0.1}{eV}$, lattice constant $a \sim \SI{5e-10}{m}$, $\abs{g}^2\sim\SI{1e-31}{eV^2 m^{3}}$. By using lattice constant for parameter $a$, it is equivalent to consider a 1D wire. The value for $\tau\sim\SI{1e-13}{s}$ is calculated in \cite{Askarpour2019} when considering only the scattering due to electron-phonon coupling. Note that $\tau$ we consider in the present work represents all the contributions to the scattering of electrons including electron-phonon interactions as mentioned in Sec.\ref{sec:model}. Therefore, we roughly estimate the order of $\tau$ as the one in \cite{Askarpour2019}. The value for $\abs{g}^2\sim \SI{1e-31}{eV^2m^3}$ is determined so that it does not contradict to the results in \cite{Askarpour2019}. Since the relaxation time for electron-phonon scattering can be evaluated with $\abs{g}^2$ as $\tau_{\mathrm{el-ph}}\sim \hbar/(\abs{g}^2 N(0) (\nneq(\omega)+1)) \sim \SI{1e-13}{s}$ at $T=\SI{300}{K}$, the value of $\abs{g}^2$ is consistent with the value of $\tau$. We estimate $\Delta N'(0)$ as $\Delta \ene_F /\ene_F^3 a^3$ and $\Delta \ene_F \sim \SI{0.1}{eV}$, which is of the order of the Zeeman gap due to the doped Mn impurities \cite{Krempasky2016}. Although the coupling to the magnetic moment of Mn induces band splitting and does not induce the asymmetry of the band according to \cite{Krempasky2016}, if one can tilt the magnetization by applying a magnetic field, the coupling induces the asymmetry of the band, $\Delta \ene_F$. $\Delta\abs{g}^2$ is estimated as $\Delta\abs{g}^2/\abs{g}^2 \sim 0.01$. 

Then
\begin{align}
    j \sim (\nneq(\omega,\Tp)-\nneq(\omega, \Te)) \times \SI{1e10}{A/m^2} \\
    q \sim (\nneq(\omega,\Tp)-\nneq(\omega, \Te)) \times \SI{1e19}{W/m^3}
\end{align}
At $\Te=\SI{300}{K}$ and $\Delta T=\Tp-\Te=\SI{0.1}{K}$, $\nneq(\omega,\Tp)-\nneq(\omega, \Te)\sim 0.001$, and $j\sim\SI{1e7}{A/m^2}, q\sim\SI{1e16}{W m^{-3}}$. This value of the electric current density is experimentally observable. As we discuss later, we believe that $\Delta T=\SI{0.1}{K}$ is also experimentally realizable. Note that the inequality \eqref{eq:condition_valid_el-ph} becomes $\Delta T/T \ll 1$ for the above parameters. 

Using the same parameters, we can also estimate the quantities defined in the previous section,
\begin{subequations}
    \begin{align}
        \lambda &\sim \qty(\frac{\hbar\omega}{\kB T})^2\nneq(\omega)(\nneq(\omega)+1)\times \SI{1e17}{W {K}^{-1} m^{-3}} \\
        p
        &\sim \qty(\frac{\hbar\omega}{\kB T})^4 \qty(\nneq(\omega)(\nneq(\omega)+1))^2 \times \SI{1e9}{WK^{-2}m^{-3}} \\
        ZT 
        &\sim \frac{\hbar\omega}{\kB T}\nneq(\omega)(\nneq(\omega)+1) \times 10^{-5}  
    \end{align} 
\end{subequations}
At $T=\SI{300}{K}$, $\lambda$ is of the order of $\SI{1e17}{W K^{-1}m^{-3}}$. This corresponds to the heat current between electrons and phonons, not to the heat current flowing spatially. A similar quantity is considered in, for example, a model called 2T-model. 2T-model describes a relaxation process of an electron-phonon coupled system after laser heating \cite{Chen2006}. In this model, a parameter for the heat current between electrons and phonons is set to $\sim \SI{1e16}{W K^{-1}m^{-3}}$ for gold.

If one can heat either electrons or phonons and make temperature difference $\Delta T\sim\SI{0.1}{K}$, the corresponding power output per volume will be $p(\Delta T)^2/4 \sim \SI{1e7}{Wm^{-3}}$ at $T=\SI{300}{K}$. This is quite large because the value corresponds to power factor$\sim\SI{1e3}{WK^{-2}m^{-1}}$ for conventional thermoelectric materials when $\nabla T\sim \SI{1}{K/cm}$.
In contrast, $ZT\sim\SI{1e-5}{}$ are very small compared to conventional thermoelectric materials. This is because the relaxation through electron-phonon interaction is so fast that the heat current between electrons and phonons is much larger than the heat current flowing spatially in the conventional thermoelectric effects. The large heat current of the proposed effect results in low efficiency, but the power output is still quite large. According to Eq.\eqref{eq:ZT}, $ZT$ increases monotonically as temperature increases at high temperatures if $\tau$ is independent of temperature. However, it should be noted that we approximate the band structure near the chemical potential in the derivation of these expressions, Eq.\eqref{eq:ZT} is valid only when the temperature is much smaller than the chemical potential, $\kB T \ll \ene_F$. If $\tau$ is due to the electron-phonon interaction, $\tau$ also depends on the temperature, and it decreases as the temperature increases (Eq.\eqref{eq:tau_estimate}). In that case $ZT$ can be estimated by using Eq.\eqref{eq:tau_estimate} as
\begin{align}
	ZT \sim \qty(\frac{\Delta \abs{g}^2}{\abs{g}^2})^2 \qty(\frac{\hbar\omega \Delta N'(0)}{N(0)})^2 \frac{\hbar\omega}{\kB T} \nneq(\omega)
\end{align}
and at high temperature $\hbar\omega \ll \kB T \ll \ene_F$ where the expression is still valid, $ZT$ is independent of the temperature.

\newcommand{\BiTe}{Bi$_2$Te$_3$}
To realize our model experimentally, it is necessary to make the temperature difference between the phonon system and the electron system. One way to make such a situation is to excite only phonons (or only electrons) in a material with light or laser. By appropriately choosing the polarization and frequency of light, it would be possible to selectively excite either phonons or electrons.  It is discussed in literature \cite{Chen2006} how the temperatures of electrons and phonons change when the system is heated by ultrashort-pulsed lasers. In this case, only electrons are excited by the laser pulse. Although the duration of the temperature difference between electrons and phonons is quite short ($\sim\SI{1}{ps}$), the maximum of the temperature difference reaches of the order of \SI{1e3}{K} according to \cite{Chen2006}.
Another idea is to use both an insulator and a metal. For example, a junction of insulator and metal with different temperatures may be one possible realization. A similar situation is discussed in \cite{Schreier2013}, and it suggests that the temperature difference between electrons and phonons can be $\sim\SI{0.1}{K}$ at steady state near an interface between a metal and a magnetic insulator. 
It may be also possible to use a topological insulator such as \BiTe, instead of a metal. In that case, the corresponding electron system is a surface state of the topological insulator. However, as discussed above, we have to break time reversal symmetry and inversion symmetry to set the spatial direction through the band structure. Therefore, in the case of \BiTe, we also have to apply a magnetic field. Although the surface state is not guaranteed in the presence of the magnetic field, if the magnetic field is small so that the splitting of the surface state bands due to the magnetic field is much smaller than the thermal energy, the proposed effect is expected to be realized. The condition on the field strength $B$ is given by $\sqrt{\hbar v^2 eB} \ll \kB T$, where $v$ is the velocity of the Dirac electron. In the case of \BiTe, $v\sim\SI{1e5}{m/s}$ and thus the condition is $B\ll\SI{100}{T}$ at $T=\SI{300}{K}$. It should be noted that, from the point of view of symmetry, thermoelectric effects similar to the proposed effect may be realized even without a magnetic field if the inversion symmetry is broken. For example, in the case of \BiTe, the inversion symmetry is broken on the surface, so there is a possibility that a thermoelectric effect due to the temperature difference between electrons and phonons can be realized without a magnetic field by a mechanism different from the proposed mechanism.

So far, we have discussed the case where phonons and electrons feel different temperatures and show that the electric current can flow.
Similarly, if there are degrees of freedom which feel different temperatures from that of electrons, then similar physics could occur. For example, one can replace phonons with another electron system. In that case, an electron density fluctuation in one system induces electric currents in the other electronic system mediated by electron-electron interactions. Another candidate is spin degrees of freedom. In the case of the magnon-driven spin Seebeck effect \cite{Xiao2010a}, the temperature difference between electrons and magnons induces a spin current. It is also possible to induce electric current by the temperature difference between the electron system and the spin system if there is a certain interaction between them, e.g., spin-orbit interaction. In any of these systems, it should break the inversion symmetry to induce finite thermoelectric effects.

In conclusion, we propose a new thermoelectric effect induced by the temperature difference between phonons and electrons. This effect can be realized in metal-insulator junctions. Similar effects are expected in noncentrosymmetric systems if one can make a temperature difference between electrons and other degrees of freedom such as spins.

\begin{acknowledgments}
The authors are grateful to S. Maekawa for insightful discussions and useful comments.
N.N. was supported by JST CREST Grant Number JPMJCR1874 and
JPMJCR16F1, Japan, and JSPS KAKENHI Grant Number 18H03676.
\end{acknowledgments}

\appendix

\begin{widetext}
\section{Solution of the Boltzmann equation} \label{Ap:sol_BE}
The solution of the Boltzmann equation Eq.\eqref{eq:Boltzmann_ss} is given by 
\begin{align}
    f_k &= \feq_k + \tau\sum_{k'}\qty(W_{k'\to k} - W_{k\to k'}) 
\end{align}
The summation over $k'$ can be explicitly calculated as follows. First we consider the summation over $k'\in \LL$ when $k\in \RR$. Then, the summation becomes 
    \begin{align}
        &\sum_{k'\in\LL}\qty(W_{k'\to k} - W_{k\to k'}) \nonumber\\
        &= \sum_{k'\in\LL} \frac{2\pi}{\hbar} \frac{\abs{\gB}^2}{V} F(\xi_{k'}, \xi_k) \qty[\delta(\ene_{k'} - \ene_k+\hbar\omega) + \delta(\ene_{k'}-\ene_k-\hbar\omega)]\delta \nneq(\omega)  \nonumber\\ 
        & \quad - \sum_{k'\in\LL} \frac{2\pi}{\hbar} \frac{\abs{\gB}^2}{V} F(\xi_k, \xi_{k'}) \qty[\delta(\ene_{k} - \ene_{k'}+\hbar\omega) + \delta(\ene_{k}-\ene_{k'}-\hbar\omega) ]\delta \nneq(\omega) \\
        &= \frac{2\pi}{\hbar} \abs{\gB}^2 \qty[F(\xi_k-\hbar\omega,\xi_k) -F(\xi_k,\xi_k-\hbar\omega)] \NL(\xi_k-\hbar\omega) \delta \nneq(\omega) \nonumber\\
        & \quad + \frac{2\pi}{\hbar} \abs{\gB}^2 \qty[F(\xi_k+\hbar\omega,\xi_k)-F(\xi_k,\xi_k+\hbar\omega)]\NL(\xi_k+\hbar\omega) \delta \nneq(\omega) 
    \end{align}     
where the definitions of $\feq$, $F$, $\delta\nneq$ are given in the main text. 

Similarly, we can take the summation for other cases as 
\begin{align}
    \text{$k\in\RR, k'\in\RR$:} \quad &\sum_{k'\in\RR}\qty(W_{k'\to k} - W_{k\to k'}) \nonumber\\
    &= \frac{2\pi}{\hbar} \abs{\gF}^2 \qty[F(\xi_k-\hbar\omega,\xi_k)-F(\xi_k,\xi_k-\hbar\omega)] \NR(\xi_k-\hbar\omega) \delta\nneq(\omega) \nonumber \\
    & \quad + \frac{2\pi}{\hbar} \abs{\gF}^2 \qty[F(\xi_k+\hbar\omega,\xi_k)-F(\xi_k,\xi_k+\hbar\omega)]\NR(\xi_k+\hbar\omega)\delta\nneq(\omega) \\
    \text{$k\in\LL, k'\in\LL$:} \quad &\sum_{k'\in\LL}\qty(W_{k'\to k} - W_{k\to k'}) \nonumber\\
    &= \frac{2\pi}{\hbar} \abs{\gF}^2 \qty[F(\xi_k-\hbar\omega,\xi_k)-F(\xi_k,\xi_k-\hbar\omega)] \NL(\xi_k-\hbar\omega) \delta\nneq(\omega) \nonumber\\
    & \quad + \frac{2\pi}{\hbar} \abs{\gF}^2 \qty[F(\xi_k+\hbar\omega,\xi_k)-F(\xi_k,\xi_k+\hbar\omega)]\NL(\xi_k+\hbar\omega)\delta\nneq(\omega) \\
    \text{$k\in\LL, k'\in\RR$:} \quad &\sum_{k'\in\RR}\qty(W_{k'\to k} - W_{k\to k'}) \nonumber\\
    &= \frac{2\pi}{\hbar} \abs{\gB}^2 \qty[F(\xi_k-\hbar\omega,\xi_k)-F(\xi_k,\xi_k-\hbar\omega)] \NR(\xi_k-\hbar\omega)\delta\nneq(\omega) \nonumber\\
    & \quad + \frac{2\pi}{\hbar} \abs{\gB}^2 \qty[F(\xi_k+\hbar\omega,\xi_k)-F(\xi_k,\xi_k+\hbar\omega)]\NR(\xi_k+\hbar\omega)\delta\nneq(\omega) 
\end{align}

Therefore, we obtain the result Eq.\eqref{eq:k_R_Wsum} and Eq.\eqref{eq:k_L_Wsum}.

\section{Calculation of the current due to the temperature difference} \label{Ap:calc_J_by_dT}
The electric current and the heat current due to the temperature difference between phonons and electrons can be calculated explicitly as follows. 

The electric current can be obtained from Eq.\eqref{eq:current} by using Eqs.\eqref{eq:k_R_Wsum},\eqref{eq:k_L_Wsum}. For example, we can take the summation over $k \in \RR$ in Eq.\eqref{eq:current} explicitly as:
\begin{align}
    &\frac{1}{L}\sum_{k\in\RR, k'} \pdv{\xi_k}{k} \qty(W_{k'\to k}-W_{k\to k'}) \\
    &= \int_\RR \frac{\dd{\xi}}{2\pi}(-1)^{\RR} \left\{\frac{2\pi}{\hbar} \qty(\abs{\gB}^2\NL(\xi-\hbar\omega) + \abs{\gF}^2\NR(\xi-\hbar\omega)) \qty[F(\xi-\hbar\omega,\xi) -F(\xi,\xi-\hbar\omega)] \right. \nonumber \\
    & \quad \left. + \frac{2\pi}{\hbar} \qty(\abs{\gB}^2\NL(\xi+\hbar\omega) + \abs{\gF}^2\NR(\xi+\hbar\omega)) \qty[F(\xi+\hbar\omega,\xi)-F(\xi,\xi+\hbar\omega)]\right\} \delta\nneq(\omega)
\end{align}
Here we used Eq.\eqref{eq:k_R_Wsum}. 
The factor $(-1)^{\RR}=\sign(\eval{\pdv{\xi}{k}}_{\RR})$ comes from the conversion of $k$ summation into $\xi$ integral,
\begin{align}
    \frac{1}{L}\sum_{k\in \RR}\pdv{\xi_k}{k} &= \int_{\RR} \frac{\dd{k}}{2\pi} \pdv{\xi_k}{k} = \int_{\RR} \frac{\dd{\xi}}{2\pi} \abs{\pdv{k}{\xi_k}}\pdv{\xi_k}{k} = \int_{\RR} \frac{\dd{\xi}}{2\pi} (-1)^\RR
\end{align}
Since the integrand is appreciable only when $\xi \simeq 0$, we can extend the range of integration to $[-\infty, \infty]$. Then we obtain 
\begin{align}
    &\frac{1}{L}\sum_{k\in\RR} \pdv{\xi_k}{k} \qty(W_{k'\to k}-W_{k\to k'}) \nonumber \\
    &= \int_{-\infty}^{\infty} \frac{\dd{\xi}}{\hbar}(-1)^{\RR} \left\{ 
    \qty(\abs{\gB}^2\NL(\xi-\hbar\omega) + \abs{\gF}^2\NR(\xi-\hbar\omega)) \qty[F(\xi-\hbar\omega,\xi)-F(\xi,\xi-\hbar\omega)]  \right. \nonumber \\
    & \quad \left. + \qty(\abs{\gB}^2\NL(\xi) + \abs{\gF}^2\NR(\xi)) \qty[F(\xi,\xi-\hbar\omega)-F(\xi-\hbar\omega,\xi)]\right\}\delta\nneq(\omega) \nonumber \\
    &= \int_{-\infty}^{\infty} \frac{\dd{\xi}}{\hbar}(-1)^{\RR} \qty(\abs{\gB}^2(\NL(\xi-\hbar\omega)-\NL(\xi)) + \abs{\gF}^2(\NR(\xi-\hbar\omega)-\NR(\xi)))\qty[F(\xi-\hbar\omega,\xi) \nneq(\omega)-F(\xi,\xi-\hbar\omega)\qty(\nneq(\omega)+1)] 
\end{align}
and similar for summation over $k\in\LL$. Therefore, we obtain 
\begin{align}
    &\frac{1}{L}\sum_{k, k'} \pdv{\xi_k}{k} \qty(W_{k'\to k}-W_{k\to k'}) \nonumber \\
    &= \int_{-\infty}^{\infty} \frac{\dd{\xi}}{\hbar}(-1)^{\RR} \qty(\abs{\gF}^2(\NL(\xi-\hbar\omega)-\NL(\xi)) + \abs{\gB}^2(\NR(\xi-\hbar\omega)-\NR(\xi))) \qty[F(\xi-\hbar\omega,\xi) \nneq(\omega)-F(\xi,\xi-\hbar\omega)\qty(\nneq(\omega)+1)] \nonumber\\
    &+ \int_{-\infty}^{\infty} \frac{\dd{\xi}}{\hbar}(-1)^{\LL} \qty(\abs{\gB}^2(\NL(\xi-\hbar\omega)-\NL(\xi)) + \abs{\gF}^2(\NR(\xi-\hbar\omega)-\NR(\xi))) \qty[F(\xi-\hbar\omega,\xi) \nneq(\omega)-F(\xi,\xi-\hbar\omega)\qty(\nneq(\omega)+1)] \nonumber \\
    &= \int_{-\infty}^{\infty} \frac{\dd{\xi}}{\hbar}(-1)^{\RR} \qty(\abs{\gF}^2-\abs{\gB}^2) \qty((\NL(\xi-\hbar\omega)-\NL(\xi)) - (\NR(\xi-\hbar\omega)-\NR(\xi))) \qty[F(\xi-\hbar\omega,\xi)-F(\xi,\xi-\hbar\omega)]\delta\nneq(\omega) \label{eq:Wsum1_again}
\end{align}
where we assumed that $(-1)^\LL = -(-1)^\RR$. From Eq.\eqref{eq:Wsum1_again}, we obtain Eq.\eqref{eq:Wsum1} in the main text. The following formulas and approximations are useful.
\begin{align}
    \int_{-\infty}^\infty \dd{\xi}F(\xi, \xi-\hbar\omega) &= \hbar\omega \nneq(\omega, \Te) \\
    \int_{-\infty}^\infty \dd{\xi}F(\xi-\hbar\omega, \xi) &= \hbar\omega(\nneq(\omega, \Te)+1) \\
    \NR(\xi-\hbar\omega)-\NR(\xi) &\simeq %
    -\hbar\omega \NR'(0) \\
    \NL(\xi-\hbar\omega)-\NL(\xi) &\simeq %
    -\hbar\omega \NL'(0)
\end{align}
where $\nneq(\omega, \Te)$ is the Bose distribution function with temperature $\Te$. By making use of these formulas and approximations, we can simplify Eq.\eqref{eq:Wsum1} as 
\begin{align}
    &\frac{1}{L}\sum_{k, k'} \pdv{\xi_k}{k} \qty(W_{k'\to k}-W_{k\to k'}) 
    = \frac{(-1)^{\RR}}{\hbar} \qty(\abs{\gF}^2-\abs{\gB}^2) (-\hbar^2\omega^2)\qty(\NL'(0)-\NR'(0)) \delta\nneq(\omega)
\end{align}
Therefore, from Eq.\eqref{eq:current}, we obtain the result Eq.\eqref{eq:current_by_T}.

Similarly, we can obtain the expression for the heat current by using Eqs.\eqref{eq:k_R_Wsum},\eqref{eq:k_L_Wsum},\eqref{eq:heat_current_def1}.
\begin{align}
    J_Q 
    &= V\int\dd{\xi}\frac{2\pi}{\hbar} \xi \qty[\NR(\xi)\qty(\abs{\gB}^2\NL(\xi-\hbar\omega) + \abs{\gF}^2\NR(\xi-\hbar\omega))+ \NL(\xi)\qty(\abs{\gF}^2\NL(\xi-\hbar\omega) + \abs{\gB}^2\NR(\xi-\hbar\omega))] \nonumber \\
    & \times \qty[F(\xi-\hbar\omega,\xi)-F(\xi,\xi-\hbar\omega)]\delta\nneq(\omega) \nonumber \\
    &+ V\int\dd{\xi}\frac{2\pi}{\hbar} \xi \qty[\NR(\xi)\qty(\abs{\gB}^2\NL(\xi+\hbar\omega) + \abs{\gF}^2\NR(\xi+\hbar\omega))+ \NL(\xi)\qty(\abs{\gF}^2\NL(\xi+\hbar\omega) + \abs{\gB}^2\NR(\xi+\hbar\omega))] \nonumber\\
    & \times \qty[F(\xi+\hbar\omega,\xi)-F(\xi,\xi+\hbar\omega)]\delta\nneq(\omega) \nonumber\\
    &\simeq 2\pi V \omega \qty[\NR(0)\qty(\abs{\gB}^2\NL(0) + \abs{\gF}^2\NR(0))+ \NL(0)\qty(\abs{\gF}^2\NL(0) + \abs{\gB}^2\NR(0))] \nonumber\\
    & \quad  \times \int\dd{\xi}\qty[F(\xi-\hbar\omega,\xi)-F(\xi,\xi-\hbar\omega)]\delta\nneq(\omega) \nonumber\\
    &= 2\pi V \hbar\omega^2 \qty[\abs{\gF}^2(\NL(0)^2+\NR(0)^2) + 2\abs{\gB}^2 \NR(0)\NL(0)]\qty[\nneq(\omega,\Tp)-\nneq(\omega, \Te)] \label{eq:heat_current_by_T_again}
\end{align}
where we have used $\sum_{k} = V\int\dd{\xi}N(\xi)$ with the volume $V$, and an approximation $N_{\mathrm{R/L}}(\xi)\simeq N_{\mathrm{R/L}}(0)$. The last expression is Eq.\eqref{eq:heat_current_by_T} in the main text. 

\section{Interpretation of the definition of the heat current}
We can transform the definition of the heat current Eq.\eqref{eq:heat_current_def1} into the following form.
\begin{align}
    J_Q &= \sum_{k,k'}(\xi_k-\xi_{k'})\Wtot_{k'\to k} \nonumber\\
    &= \sum_{k,k'} \frac{2\pi}{\hbar} \frac{\abs{g_{k'-k}}^2}{V} f_{k'}\qty(1-f_k)\hbar\omega_{k'-k} \qty[n_{k'-k} \delta(\ene_{k'} - \ene_k+\hbar\omega_{k'-k}) - \qty(n_{k'-k}+1)\delta(\ene_{k'}-\ene_k-\hbar\omega_{k'-k})] \label{eq:heat_current_def2}
\end{align}
Here, we have used $\Wtot_{k\to k'}$ instead of $W_{k\to k'}$. The form of Eq.\eqref{eq:heat_current_def2} might be easier to intuitively understand than Eq.\eqref{eq:heat_current_def1}. The first term in the summation in Eq.\eqref{eq:heat_current_def2} represents the probability for electrons to absorb phonon multiplied by the absorbed energy $\hbar\omega_{k'-k}$. Similarly, the second term corresponds to the phonon-emitting process.

\section{Calculation of $L_{11}$ and $L_{21}$} \label{Ap:calc_linres}
Here we calculate $L_{11}$ and $L_{21}$. What we want to calculate is the electric current $J$ and heat current $J_Q$ in the presence of an electric field $E$ up to $\order{E\abs{g_q}^2}$. By solving the Boltzmann equation up to $\order{E\abs{g_q}^2}$, we obtain
\begin{align}
    f_k &= \feq_k + f_k^{(1)} \\
    f_k^{(1)} &= -\frac{eE\tau}{\hbar} \pdv{\feq_k}{k} \label{eq:Fk1}
\end{align}
Here, we have used Eq.\eqref{eq:Boltzmann_ss} based on the assumption that the scattering term proportional to $f_k-\feq_k$ is incorporated into the relaxation time term.

Using $f_k^{(1)}$ in Eq.\eqref{eq:Fk1}, the currents can be written as follows.
\begin{align}
    J &= \frac{1}{L}\sum_{k} \frac{e}{\hbar}\pdv{\xi_k}{k}f_k^{(1)} \\
    J_Q &= \sum_{k} \xi_k \qty(\pdv{f_k}{t})_{\mathrm{phonon}} 
    = \sum_{k}\xi_k\sum_{k'}\qty(\Wtot_{k'\to k}-\Wtot_{k\to k'}) 
    = \sum_{k,k'}(\xi_k-\xi_{k'})\Wtot_{k'\to k} \nonumber \\
    &= \sum_{k,k'}(\xi_k-\xi_{k'})\frac{2\pi}{\hbar}\frac{\abs{g_{k'-k}}^2}{V} \qty(f_{k'}^{(1)}(1-\feq_k) - \feq_{k'}f_k^{(1)}) 
    \left[n(\omega_{k'-k})\delta(\xi_{k'}-\xi_k+\hbar\omega_{k'-k}) \right. 
    \left. + (n(\omega_{k'-k})+1)\delta(\xi_k-\xi_{k'} + \hbar\omega_{k'-k}) \right] \label{eq:JQ_L21_calc}
\end{align}
Note that $\Wtot_{k'\to k}$ instead of $W_{k'\to k}$ is used in the calculation of $J_Q$.
Below, we calculate these quantities using the simplification explained in the main text, i.e., $\omega_{k'-k} = \omega$ and the assumption for $g_{k'-k}$, Eq.\eqref{eq:gq_dependence_assumption}.

First, let us calculate $J$. This is the well-known contribution to the electric current and can be calculated as follows. 
\begin{align}
    J &= \frac{1}{L}\sum_{\alpha=\RR,\LL}\sum_{k\in\alpha}\frac{e}{\hbar}\pdv{\xi_k}{k} f_k^{(1)} 
    = -\frac{1}{L}\frac{e^2E\tau}{\hbar^2}\sum_{\alpha=\RR,\LL}\sum_{k\in\alpha}\qty(\pdv{\xi_k}{k})^2 \pdv{\feq_k}{\xi_k} \nonumber \\
    &= -\frac{e^2E\tau}{\hbar^2}\sum_{\alpha=\RR,\LL}\int_{\alpha}\frac{\dd{k}}{2\pi}\qty(\pdv{\xi_k}{k})^2 \pdv{\feq_k}{\xi_k} 
    = -\frac{e^2E\tau}{\hbar^2}\sum_{\alpha=\RR,\LL}\int_{\alpha} a^2N_\alpha(\xi)\dd{\xi} \qty(\frac{1}{2\pi a^2N_{\alpha}(\xi)})^2 \pdv{\feq_k}{\xi_k} \nonumber \\
    &= -\frac{e^2E\tau}{4\pi^2 a^2 \hbar^2}\sum_{\alpha=\RR,\LL}\int_{\alpha}\dd{\xi} \frac{1}{N_{\alpha}(\xi)} \pdv{\feq_k}{\xi_k} 
\end{align}
Here we used a relation $N_{\alpha}(\xi) = \qty(2\pi a^2\abs{\pdv{\xi_k}{k}})^{-1}$. As in the calculation of the currents in Appendix \ref{Ap:calc_J_by_dT}, we neglect the $\xi$ dependence of $N_{\alpha}(\xi)$ and approximate it by $N_{\alpha}(\xi=0)$. Then we obtain 
\begin{align}
    J &= -\frac{e^2E\tau}{4\pi^2a^2\hbar^2}\sum_{\alpha=\RR,\LL}\int_{\alpha}\dd{\xi} \frac{1}{N_{\alpha}(0)} \pdv{\feq_k}{\xi_k} = \frac{e^2E\tau}{4\pi^2a^2\hbar^2}\qty(\frac{1}{\NR(0)}+\frac{1}{\NL(0)}) = \sigma_0 E \label{eq:J1_E_calc_results}
\end{align}

Similary, we can calculate $L_{21}$. Starting from \eqref{eq:JQ_L21_calc}, 
\begin{align}
    &J_Q %
    = \sum_{\alpha,\beta\in\RR,\LL}\sum_{k\in\alpha,k'\in\beta}(\xi_k-\xi_{k'})\frac{2\pi}{\hbar}\frac{\abs{g_{\beta\alpha}}^2}{V} \qty(f_{k'}^{(1)}(1-\feq_k) - \feq_{k'}f_k^{(1)}) \qty(n(\omega)\delta(\xi_{k'}-\xi_k+\hbar\omega) + (n(\omega)+1)\delta(\xi_k-\xi_{k'} + \hbar\omega)) \nonumber \\
    &= \sum_{\alpha,\beta\in\RR,\LL}\sum_{k\in\alpha,k'\in\beta}E\frac{2\pi e\omega\tau}{\hbar}\frac{\abs{g_{\beta\alpha}}^2}{V} \qty[- \pdv{\feq_{k'}}{k'} + \pdv{\feq_{k'}}{k'}\feq_k + \feq_{k'} \pdv{\feq_k}{k}] \qty(n(\omega)\delta(\xi_{k'}-\xi_k+\hbar\omega) - (n(\omega)+1)\delta(\xi_k-\xi_{k'} + \hbar\omega)) \label{eq:JQ_E_calc_intermediate1}
\end{align}
The first term in the square bracket in \eqref{eq:JQ_E_calc_intermediate1} reads 
\begin{align}
    &\sum_{\alpha,\beta\in\RR,\LL}\sum_{k\in\alpha,k'\in\beta} E\frac{2\pi e\omega\tau}{\hbar}\frac{\abs{g_{\beta\alpha}}^2}{V} \qty(- \pdv{\feq_{k'}}{k'}) \qty[n(\omega)\delta(\xi_{k'}-\xi_k+\hbar\omega) - (n(\omega)+1)\delta(\xi_k-\xi_{k'} + \hbar\omega)] \nonumber \\
    &= \sum_{\alpha,\beta\in\RR,\LL}\sum_{k'\in\beta} E\frac{2\pi e\omega\tau}{\hbar}\abs{g_{\beta\alpha}}^2 \qty(- \pdv{\feq_{k'}}{k'}) \qty[n(\omega)N_{\alpha}(\xi_{k'}+\hbar\omega) - (n(\omega)+1)N_{\alpha}(\xi_{k'}-\hbar\omega)]
\end{align}
The second and the third terms in the square bracket in \eqref{eq:JQ_E_calc_intermediate1} read 
\begin{align}
    &\sum_{\alpha,\beta\in\RR,\LL}\sum_{k\in\alpha,k'\in\beta}E\frac{2\pi e\omega\tau}{\hbar}\frac{\abs{g_{\beta\alpha}}^2}{V} \qty[\pdv{\feq_{k'}}{k'}\feq_k + \feq_{k'} \pdv{\feq_k}{k}] \qty(n(\omega)\delta(\xi_{k'}-\xi_k+\hbar\omega) - (n(\omega)+1)\delta(\xi_k-\xi_{k'} + \hbar\omega)) \nonumber \\
    &= -\sum_{\alpha,\beta\in\RR,\LL}\sum_{k\in\alpha,k'\in\beta}E\frac{2\pi e\omega\tau}{\hbar}\frac{\abs{g_{\beta\alpha}}^2}{V}\pdv{\feq_{k'}}{k'}\feq_k \qty(\delta(\xi_{k'}-\xi_k+\hbar\omega) + \delta(\xi_k-\xi_{k'} + \hbar\omega)) \nonumber \\
    &= -\sum_{\alpha,\beta\in\RR,\LL}\sum_{k'\in\beta}E\frac{2\pi e\omega\tau}{\hbar}\abs{g_{\beta\alpha}}^2 \pdv{\feq_{k'}}{k'} \qty(\feq(\xi_k' + \hbar\omega)N_{\alpha}(\xi_{k'}+\hbar\omega) + \feq(\xi_k'-\hbar\omega)N_{\alpha}(\xi_{k'}-\hbar\omega))
\end{align}
Therefore, $J_Q$ is 
\begin{align}
    J_Q &= -E\frac{2\pi e\omega\tau}{\hbar}\sum_{\alpha,\beta\in\RR,\LL}\sum_{k'\in\beta} \left\{\abs{g_{\beta\alpha}}^2 \pdv{\feq_{k'}}{k'} \qty[n(\omega)N_{\alpha}(\xi_{k'}+\hbar\omega) - (n(\omega)+1)N_{\alpha}(\xi_{k'}-\hbar\omega)] \right. \nonumber \\
    & \left. + \abs{g_{\beta\alpha}}^2 \pdv{\feq_{k'}}{k'} \qty(\feq(\xi_{k'} + \hbar\omega)N_{\alpha}(\xi_{k'}+\hbar\omega) + \feq(\xi_{k'}-\hbar\omega)N_{\alpha}(\xi_{k'}-\hbar\omega))\right\} \nonumber \\
    &= -E\frac{2\pi e\omega\tau L}{\hbar}\sum_{\alpha,\beta\in\RR,\LL}\abs{g_{\beta\alpha}}^2\int_{\beta}\frac{\dd{\xi_{\beta}}}{2\pi}(-1)^{\beta}  \qty[{\feq}'(\xi_{\beta})\feq(\xi_\beta + \hbar\omega)N_{\alpha}(\xi_{\beta}+\hbar\omega) + {\feq}'(\xi_{\beta}+\hbar\omega)(\feq(\xi_{\beta})-1)N_{\alpha}(\xi_{\beta})] \nonumber\\
    & -E\frac{2\pi e\omega\tau L}{\hbar}\sum_{\alpha,\beta\in\RR,\LL}\abs{g_{\beta\alpha}}^2\int_{\beta}\frac{\dd{\xi_{\beta}}}{2\pi}(-1)^{\beta} {\feq}'(\xi_{\beta}) n(\omega)\qty[N_{\alpha}(\xi_{\beta}+\hbar\omega)-N_{\alpha}(\xi_{\beta}-\hbar\omega)] \nonumber \\
    &= -E\frac{2\pi e\omega\tau L}{\hbar}\sum_{\alpha,\beta\in\RR,\LL}\abs{g_{\beta\alpha}}^2\int_{\beta}\frac{\dd{\xi_{\beta}}}{2\pi}(-1)^{\beta}  [{\feq}'(\xi_{\beta})\feq(\xi_\beta + \hbar\omega)N_{\alpha}(\xi_{\beta}+\hbar\omega) - {\feq}(\xi_{\beta}+\hbar\omega){\feq}'(\xi_{\beta})N_{\alpha}(\xi_{\beta}) \nonumber\\
    &-{\feq}(\xi_{\beta}+\hbar\omega)(\feq(\xi_{\beta})-1)N_{\alpha}'(\xi_{\beta}) + {\feq}'(\xi_{\beta}) n(\omega)\qty(N_{\alpha}(\xi_{\beta}+\hbar\omega)-N_{\alpha}(\xi_{\beta}-\hbar\omega))] \quad \text{($f'(f-1)N$ is partially integrated)} \nonumber \\
    &= -E\frac{2\pi e\omega\tau L}{\hbar}\sum_{\alpha,\beta\in\RR,\LL}\abs{g_{\beta\alpha}}^2\int_{\beta}\frac{\dd{\xi_{\beta}}}{2\pi}(-1)^{\beta} \qty[ {\feq}'(\xi_{\beta})\feq(\xi_\beta + \hbar\omega)\qty[N_{\alpha}(\xi_{\beta}+\hbar\omega) - N_{\alpha}(\xi_{\beta})] + {\feq}(\xi_{\beta}+\hbar\omega)(1-\feq(\xi_{\beta}))N_{\alpha}'(\xi_{\beta})] \nonumber \\
    & -E\frac{2\pi e\omega\tau L}{\hbar}\sum_{\alpha,\beta\in\RR,\LL}\abs{g_{\beta\alpha}}^2\int_{\beta}\frac{\dd{\xi_{\beta}}}{2\pi}(-1)^{\beta} {\feq}'(\xi_{\beta}) \qty[n(\omega)(N_{\alpha}(\xi_{\beta}+\hbar\omega)-N_{\alpha}(\xi_{\beta}-\hbar\omega))] 
\end{align} 
If we neglect $\xi$-dependence of $N_{\alpha}$ completely, $J_Q$ becomes 0. Therefore, we expand $N_{\alpha}(\xi_\beta \pm \hbar\omega)$ up to $\order{\hbar\omega}$ and approximate it as $N_{\alpha}(\xi_\beta \pm \hbar\omega)-N_{\alpha}(\xi_\beta) \simeq\pm \hbar\omega N_{\alpha}'(0)$. Then we obtain
\begin{align}
    J_Q %
    &\simeq -E\frac{2\pi e\omega\tau L}{\hbar}\sum_{\alpha,\beta\in\RR,\LL}\abs{g_{\beta\alpha}}^2 N_{\alpha}'(0)\int_{\beta}\frac{\dd{\xi_{\beta}}}{2\pi}(-1)^{\beta}  [\hbar\omega{\feq}'(\xi_{\beta})\feq(\xi_\beta + \hbar\omega) + F(\xi_{\beta}+\hbar\omega,\xi_{\beta})] \nonumber \\
    & -E\frac{2\pi e\omega\tau L}{\hbar}\sum_{\alpha,\beta\in\RR,\LL}\abs{g_{\beta\alpha}}^2 2\hbar\omega N_{\alpha}'(0)\int_{\beta}\frac{\dd{\xi_{\beta}}}{2\pi}(-1)^{\beta} {\feq}'(\xi_{\beta}) n(\omega) \nonumber \\
    &= -E\frac{2\pi e\omega\tau L}{\hbar}\sum_{\alpha,\beta\in\RR,\LL}\abs{g_{\beta\alpha}}^2 \hbar\omega N_{\alpha}'(0)\frac{(-1)^{\beta}}{2\pi} ( - \beta\hbar\omega\nneq(\omega,\Te)(\nneq(\omega,\Te)+1)) \nonumber\\
    &= Ee\beta\hbar\omega^3\tau L \nneq(\omega,\Te)(\nneq(\omega,\Te)+1)\sum_{\alpha,\beta\in\RR,\LL}(-1)^{\beta} \abs{g_{\beta\alpha}}^2  N_{\alpha}'(0) \nonumber\\
    &= E (-1)^\RR e\beta\hbar\omega^3\tau L (\abs{\gF}^2-\abs{\gB}^2)(\NR'(0)-\NL'(0))\nneq(\omega,\Te)(\nneq(\omega,\Te)+1) \label{eq:derive_JQ_by_E}
\end{align}
where we have assumed again that $(-1)^\RR = -(-1)^\LL$. From Eq.\eqref{eq:derive_JQ_by_E}, the expression for $L_{21}$, Eq.\eqref{eq:L21}, is obtained.
\end{widetext}

\bibliographystyle{apsrev4-2}
\bibliography{phonon_ratchet_ref}

\end{document}